# Intention to Use SMS Vaccination Reminder and Management System among Health Centers in Malaysia: The Mediating Effect of Attitude

*Kamal Karkonasasi[1], Cheah Yu-N[2], Seyed Aliakbar Mousavi[3]

School of Computer Sciences, Universiti Sains Malaysia (USM), Penang, 11800, Malaysia

Email: asasi.kamal@gmail.com[1], yncheah@usm.my [2], pouyaye@gmail.com[3],

*corresponding Author

## ABSTRACT

### Background

The majority of health centers in Malaysia still use a paper-based vaccination system to keep a record of the vaccination uptakes of young children and, to remind their parents, which causes workload on nurses. Moreover, it is not a cost-effective procedure. In order to overcome the disadvantages of the current procedure, we propose a system called Virtual Health Connect. It is a SMS vaccination reminder and management system. The purpose of this study is to find out influential factors of intention to use the proposed system among health centers by considering the technology acceptance model as the underlying theory.

### Methods

The proposed model of this study is tested using data collected from self-directed questionnaires filled up by 70 nurses in government hospitals and government clinics in Malaysia. A research method, which is based on the multi-analytical approach of Multiple Regression Analysis and Artificial Neural Networks, helps to refine the results of this study.

### Results

The compatibility of the proposed system only has a significant and positive effect on attitude to use the system among other factors. Moreover, health centers' intention to use the system is further influenced by their perceived usefulness than their attitude to use. The mediating effect of Attitude is also proved. However, there is not any statistically significant difference in Intention scores accounting for the participants' demographic characteristics.

### Conclusions

The implication of this study is that by developing the system, which fits to the current procedures of health centers, we give them positive feelings. IT educational programs also should be provided for nurses to identify the benefits of the proposed system. It is also important to consider the indirect effects from independent variables on intention to use the system. However, the system does not require to consider the demographic characteristics of its end users.

## 1. Introduction

Vaccination is one of the most fruitful and cost-effective public health interventions (Domek et al., 2016). However, the number of young children with overdue or missed vaccination uptakes increases due to growth in the number of working parents (Kharbanda et al., 2009) (Ooi and Cheah, 2011). In order to assure that young children are vaccinated on time, the health centers' nurses are required to send several reminders to their parents.

Currently, health centers use a paper-based vaccination system to keep track of the vaccination and to remind their parents through a phone call or normal mail. The current system causes workload on nurses. Moreover, the system is not cost-effective because of imposing costs through phone calls or normal mails (Ooi and Cheah, 2011).

In order to address the issues of young children with overdue or missed vaccination uptakes and to overcome the disadvantages of the current system, we propose Virtual Health Connect (VHC). VHC is a web-based system that helps health centers' nurses to record and keep track of vaccination uptakes automatically. Moreover, it sends short message service (SMS) reminders to parents to bring their young children for vaccination uptakes.

SMS has remarkable and unused potential for disease management in low- and middle-income countries (LMICs) (Domek et al., 2016). A SMS vaccination reminder is a useful and effective system in order to save time, labour and cost of managing vaccination records (Ooi and Cheah, 2011; Domek et al., 2016). The effectiveness of SMS vaccination reminders has also been proved for different types of vaccinations around the world (Domek et al., 2016; Ooi and Cheah, 2011; Kharbanda et al., 2011; Stockwell et al., 2012; Wakadha et al., 2013; Kaewkungwal et al., 2010; Vilella et al., 2004; Abbas and Yusof, 2011; Harvey et al., 2015; Jacobson and Szilagyi, 2005). The penetration rate for the mobile phone subscriptions in Malaysia in the year 2014 was 144.2 % (MCMC, 2014). It shows that parents own at least one mobile phone ready to receive SMS. Therefore, we are able to improve the vaccination coverage among young children through using VHC.

During our pilot study, 19 managers of health centers in Malaysia were asked if they have already applied a SMS vaccination reminder and management system. The majority of them responded that they have not applied it. This fact is in accordance with the studies conducted by Ooi and Cheah (2011) which showed that the level of adoption of the system among health centers in Malaysia is low.

The majorities of studies about SMS reminder system were performed in the developed countries (Cole-Lewis and Kershaw, 2010). Therefore, there are deficiencies in the literature to evaluate this system in LMIC (Gurman et al., 2012; Fjeldsoe et al., 2009; Domek et al., 2016). In response to the low-level adoption and deficiencies in the literature, this study aims to find out what factors influence intention to use VHC among health centers in Malaysia through the mediating effect of attitude. By having the technology acceptance model (TAM) as the underlying theory, the objectives of this study are to examine: 1) the relationship between Attitude and the independent variables, i.e., perceived usefulness (UseF), perceived ease of use (EaseOfUse), privacy concern (PrivConc), compatibility (Comp), and accessibility (Access). 2) the relationship between Intention and the independent variables, i.e., UseF, and Attitude. 3) the mediating effect of Attitude 4) to determine whether there are any statistically significant difference in Intention scores accounting for the participants' demographic characteristics.

The current study will contribute to the theoretical body of knowledge by extending the TAM model by adding new variables. The practical contribution of this study is to help to software developers to develop VHC by including significant variables that drive health centers to intend to use VHC directly or indirectly.

In the following sections, the recent studies about Electronic Health Record (EHR) systems using TAM based models are reviewed. In section 3, a research model is proposed. In section 4, each variable is defined, and hypotheses are proposed. In section 5, the research methodology is discussed. In section 6, the collected data from the questionnaires are analyzed. In section 7, we discuss the results. In section 8, the limitation of the current study and also the future work are stated. And finally, in the last section, a conclusion is performed.

## 2. The Recent Studies about Electronic Health Record (EHR) Using TAM Based Models

An Electronic Health Record (EHR) is a source of electronically preserved and stored information about a patient's lifelong health status and health care. Therefore, it can deliver an information management system to supply medical reminders and alerts, associations with knowledge sources for healthcare decision support, and exploration of mass data both for research and care management (Paul and Clement, 2006). SMS vaccination and reminder system is a subdomain of EHR.

To the best of our knowledge, no study looked at the enablers and inhibitors of intention to use the SMS vaccination and reminder system among health centers. Therefore, we extended our literature review about the recent studies about EHR systems using TAM based models. These studies are reviewed below.

### 2.1 Clinical Reminder System (CRS)
Johnson et al. (2014) proposed Clinical Reminder System (CRS) to be used by medicine residents at the Western Pennsylvania Hospital, USA. The residents used the system to document and recall patient care data and to create unique patient reminders to advance the management of main chronic situations and preventive care measures. They included more factors in the acceptance model of the CRS. These new factors are an individual's general optimism (GO), Computer knowledge (CK), Computer experience (CE), Self-reported usage (SRU), Initial usage (IniU), Average usage (AU) and User satisfaction (SAT). In Table 1, the new factors are explained.

Table 1, The definition of the new factors of CRS (Johnson et al., 2014)

| Factor | Definition |
|---|---|
| An individual's general optimism (GO) | It evaluates a one's general view about the type of the technology being presented instead of the assessment of a specific system or product. |
| Computer knowledge (CK) and Computer experience (CE) | These factors evaluate computer skill of CRS's end users. |
| Self-reported usage (SRU) | SRU studies the difference between the actual and self-reported usage of CRS. |
| Initial usage (IniU) and average usage (AU) | IniU are acquired one month after technology implementation. |
| User satisfaction (SAT) | SAT measures the non-usage related outcome. |

The results of the study are presented in Figure 1. GO has a significant and positive effect on Institutionalized use (IU), Usage trajectory group (UTG), and Average usage (AU). Moreover, EaseOfUse is a significant and positive predictor of SRU and SAT. However, CK has a significant and negative effect on UTG, AU, and SAT. The authors exclude Attitude, Intention, and UseF from the model.

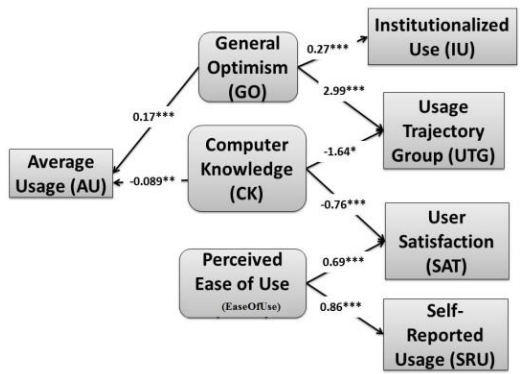

**Fig. 1**, Results of model testing (*P < .05; **P < .01; ***P < .001)
(Johnson M. P. et al., 2014)

**2.2 Electronic Health Record (EHR), First Study**
Marie-Pierre G. et al. (2014) recognized the principal factors of physician acceptance of EHR among 150 general practitioners and specialists of the province of Quebec, Canada. The authors tested four research models, i.e., TAM, Extended TAM, Psychosocial Model, and Integrated Model. The original TAM (Model 1) is shown in Figure 2. UseF and EaseOfUse have a direct and positive effect on Intention. Moreover, EaseOfUse has a direct and positive influence on UseF. This model describes 44.0% of the variance in Intention. The authors excluded Attitude from the model.

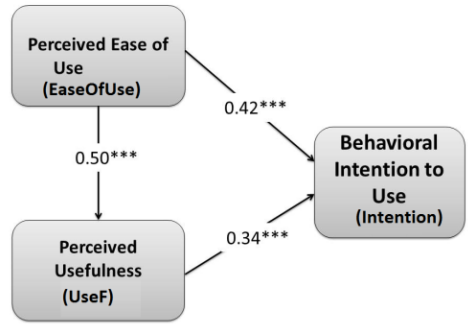

**Fig. 2**, Results of TAM (***P < 0.001)
(Marie-Pierre G et al., 2014)

The second model is extended TAM (Figure 3). More variables are added to the original TAM to make it more applicable. DR clarifies how much the impact of usage of the EHR is obvious for the general practitioners and specialists. UseF and EaseOfUse have a direct and positive impact on Intention. The researchers excluded Attitude. The model describes 44.0% of the variance in Intention. In addition, DR and EaseOfUse have a direct and positive impact on UseF, and 48.1% of the variance in this variable is described. Moreover, CSE has a direct and positive impact on EaseOfUse. CSE describes 6% of the variance in EaseOfUse. All indirect effects are also significant.

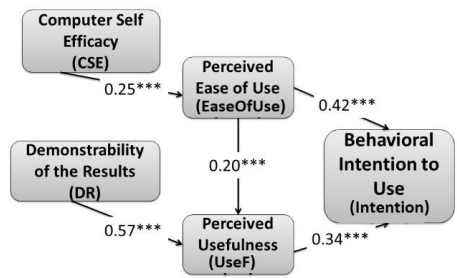

**Fig. 3**, Results of extended TAM (***P < 0.001)
(Marie-Pierre G et al., 2014)

The third model is a psychosocial model (Figure 4). It contains theoretical variables proposed by Triandis (1980), Orruno et al. (2011), Godin et al. (2008), Gagnon et al. (2012) and Gagnon et al. (2006). Personal Identity (PI) shows the attitude of the respondents toward information and communication technology (ICT). In Social Norm (SN), the respondents look mostly at the attitude toward the EHR among health professionals outside their specialty area. Whereas, in Professional Norm (PN), the respondents look at the attitude among health professionals inside their specialty area. There is no significant relationship between UseF, PI and CSE, and Intention. However, EaseOfUse, SN, and PN have a significant and positive impact on Intention. The model explains 53% of the variance in Intention.

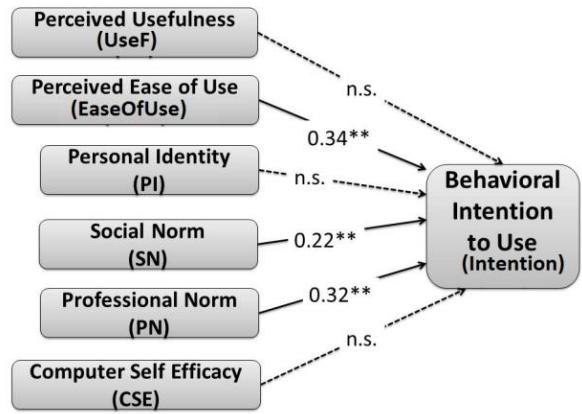

**Fig. 4**, Results of Psychosocial model (**P <= .01)
(Marie-Pierre G et al., 2014)

The last model is a combination of the TAM and the psychosocial model (Figure 5). The model includes theoretical variables that have been proposed in latest studies on physician acceptance of information technologies. These variables are DR, Resistance to change (RC), and CSE. In RC, respondents' resistance against the EHR is measured. Information about the change (IC) is defined as the level of information that respondents obtained regarding the possible changes that EHR causes in their routine works. DR, EaseOfUse, SN, and PN have a significant and positive effect on Intention. This model explains 55 % of the variance in Intention.

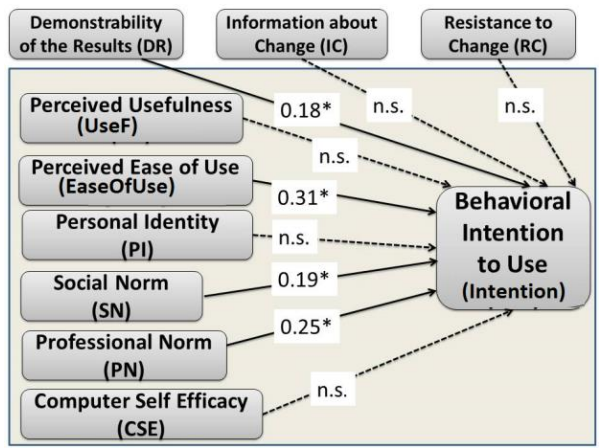

**Fig. 5**, Results of integrated model (*P < 0.05)
(Marie-Pierre G et al., 2014)

**2.3 e-Health Companion**
Elavarasen (2010) proposed e-Health Companion. This system observes the users' health for lifestyle sicknesses such as diabetes and hypertension. The biomedical instruments transfer the information to the central computer server. The server will go through the given information and will alarm the users if there is any crucial condition. A study was conducted among 60 potential users to find out what factors affect the acceptance of the system. The proposed model is TAM that is extended by several factors. These factors are Compatibility (Comp), Risk and Access. Access is defined as an individual's opinion of how easy to access to the system. Moreover, Comp is the degree to which an individual believes that the system is consistent with his/her current values, former experience, and wishes. Lastly, Risk measures the degree to which an individual sees certain negative outcomes from using the system. The proposed model is shown in Figure 6.

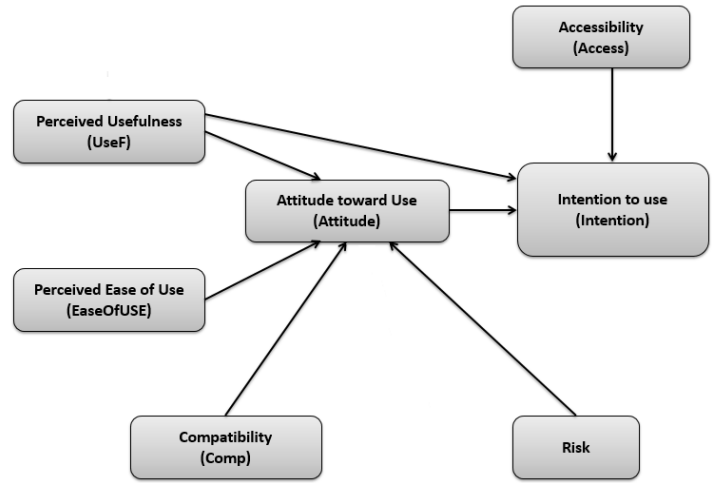

**Fig. 6**, Proposed e-Health Companion acceptance model
(Elavarasen, 2010).

The data analysis part of the study was questionable. Therefore, we only look at the proposed acceptance model and we do not go through the data analysis part.

**2.4 Electronic Health Record (EHR), Second Study**
Marie-Pierre G., et al (2016) proposed a multilevel model to discover organization-level and individual-level influential factors on EHR adoption by physicians. A prospective cross-sectional study was conducted among physicians in 49 primary healthcare organizations in four regions of the province of Quebec, Canada. 278 completed questionnaires were back from the 31 organizations that had at least five respondents. The multilevel modeling found no significant overall effect of organizational level's factors on physician intention. However, all individual-level factors have a positive and significant effect on intention. This individual-level model is shown in Figure 7 and describes 64% of the variance in Intention. The authors excluded the mediation effect of Attitude.

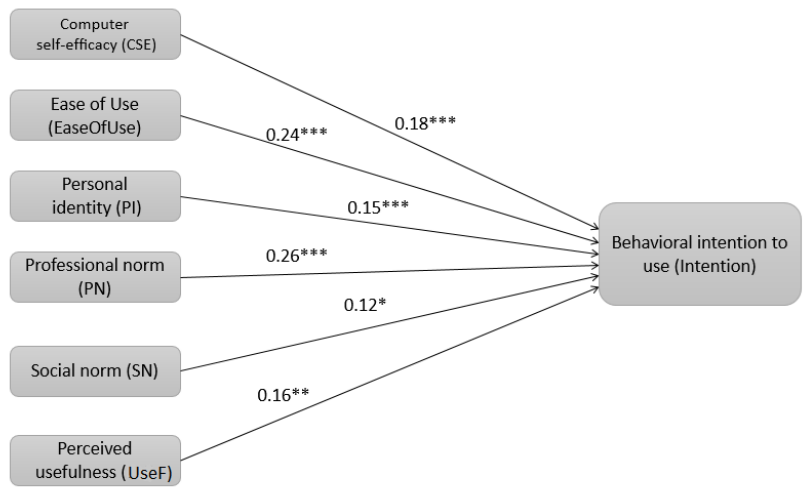

**Fig. 7**, Results of the individual-level model ($*P < 0.05$; $**P < 0.01$; $***P < 0.001$)
(Marie-Pierre G., et al. 2016)

**2.5 Electronic Medical Record (EMR)**

Aldosari et al. (2018) conducted a study to understand the perspective of the nursing staff working in Imam Abdulrahman Al Faisal Hospital, Saudi Arabia using electronic medical record (EMR) in their clinical practice. The findings show that there is a strong positive correlation between UseF and EaseOfUse resulting in a positive effect on nurses' acceptance of EMR. The findings also represent that user characteristics, system quality, and top management and IT support are important in predicting nursing staff's acceptance toward EMR. The research model is represented in Figure 14.

The concern with their study is that the authors did not examine the relationships using multiple regression analysis and merely were satisfied with Spearman's Correlation analysis, which is not a powerful analysis for hypothesis testing. Moreover, they did not perform any test to evaluate validity and reliability of their measurement items.

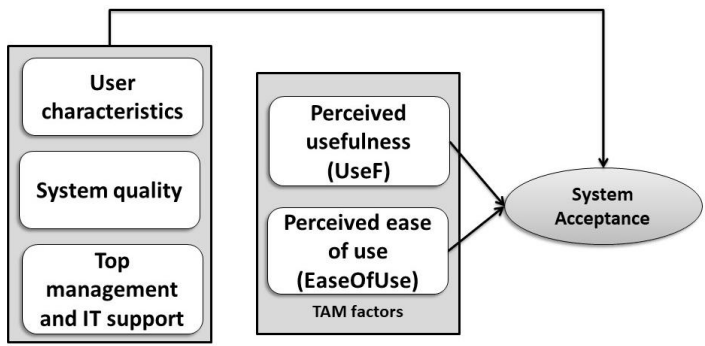

**Fig. 14**, Research model of electronic medical record (EMR)
(Aldosari et al., 2018)

## 3. The Proposed Research Model

TAM is proposed by Davis (1989) as an analytical model to predict user acceptance to use computer technology. The model is one of the most extensively used and fit models to examine the behavioral and social intentions that influence the acceptance of technology (Wu et al., 2008; Aldosari et al., 2018). TAM is the most promising model in order to predict the factors that influence the acceptance of technology across many contexts such as healthcare (Chow et al., 2012; Vathanophas and Pacharapha, 2010; Aldosari et al., 2018). In the context of this study, a TAM based model is proposed to understand what factors influence intention to use VHC through the mediating effect of Attitude. The model is shown in Figure 8.

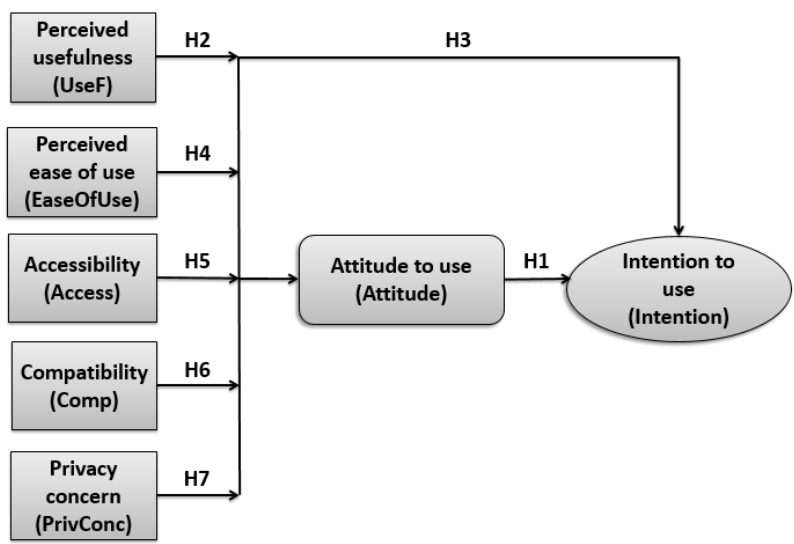

**Fig. 8**, The Proposed Research Model
*Note. H1 to H7 indicate the hypotheses. H8 to H12 (The mediating effect of Attitude)*

## 4. Proposed Hypotheses

In the following section, each factor is discussed in detail and its effect is hypothesized.

### 4.1 The Relationship between Attitude and Intention

Attitude toward using a system is defined as the individual's positive or negative feelings about working with the system, and intention to use a system is defined as the degree of the determination of the individual's aim to use the system (Davis et al., 1989, p. 984). Ajzen and Fishbein (1980) indicated that attitude influences behavioral intention. Davis (1989) also stated the significant and positive effect of Attitude on Intention. Moreover, Owitlawakul et al. (2014) indicated that attitude toward using the Electronic Health Records for Nursing Education (EHRNE) software program was the significant factor on intention to use the EHRNE. Kijsanayotin et al. (2009) also stated that Attitude influenced the intention to use medical IT. Kim et al. (2015) also stated that Attitude positively influenced the intention to use a mobile electronic medical record (MEMR). Hsiao and Chen (2015) also stated that attitudes toward using computerized clinical practice guidelines (CCPG) was critical factor influencing physicians' intention to use CCPG.

Therefore, the following hypothesis is proposed about VHC:

**H1:** Attitude will have a significant and positive effect on intention.

### 4.2 The Relationship between UseF and Attitude

UseF refers to the degree to an individual's faith that using a specific system will improve his or her job routine (Davis et al., 1989, p. 984). UseF constantly appears to be an important predictor of Attitude (Venkatesh et al., 2003). For instance, Aldosari et al. (2018) stated that UseF had a positive effect on nurses' acceptance toward using EMR. Elavarasen (2010) also claimed that UseF has a positive impact on Attitude toward using e-Health Companion. Kim et al. (2015) stated that PE positively influenced the attitude to use a MEMR. PE has similar concept with UseF.

Therefore, we claim that if the end users of VHC believe that the system helps them to manage their daily routine, they have positive feelings to use it. The following hypothesis is proposed to test our claim:

**H2:** The perceived usefulness (UseF) will have a significant and positive effect on Attitude.

### 4.3 The Relationship between UseF and Intention

Previous models on TAM stated that UseF has a positive impact on Intention, both directly and indirectly (Davis, 1989; Venkatesh and Davis, 2000; Venkatesh et al., 2008). Moreover, TAM and extended TAM models by Marie-Pierre G et al. (2014) stated UseF as a positive predictor of Intention to use EHR. Marie-Pierre G et al (2016) also supports this statement. Elavarasen (2010) also claimed that UseF has a positive impact on intention to use e-Health Companion. Aldosari et al. (2018) also stated that Usef had a positive effect on nurses' acceptance toward using EMR. Dünnebeil et al (2012) also indicated that UseF positively affected behavioral intention to use Electronic Health Services (EHS). Wahyuni (2017) also pointed out that UseF positively affected intention to use e-health services. Kuo et al. (2013) also stated that the Usef of mobile electronic medical record system (MEMRS) with nurses' work practices enhances nurses' intention to use MEMRS. Aldosari (2012) also proved that UseF of a picture archiving and communication system (PACS) positively affected intention to use PACS by staff. Dunnebeil et al. (2012) also found out that UseF positively influenced the intention to use e-health applications. Esmaeilzadeh et al. (2015) examined the intention to use a clinical decision support system (CDSS) and stated Performance Expectancy (PE) as an influential factor. Kijsanayotin et al. (2009) also stated that PE influenced the intention to use medical IT. Hsiao and Chen (2015) also stated that UseF of CCPG was critical factor influencing physicians' intention to use CCPG. However, integrated model and psychosocial model by Marie-Pierre G et al. (2014) do not show any significant impact between two variables.

Therefore, this study proposes the following hypotheses in order to clarify the relationship:

**H3:** The perceived usefulness (UseF) will have a significant and positive effect on Intention.

**4.4 The Relationship between EaseOfUse and Attitude**
Davis et al. (1989, p. 984) define EaseOfUse as a degree to an individual's faith that using a particular system would be free of struggle. Marie-Pierre G et al. (2014) and Marie-Pierre G et al. (2016) stated that EaseOfUse has a significant, positive effect on Intention to use EHR while the authors excluded Attitude from the model. Moreover, Johnson et al. (2014) stated that EaseOfUse is a significant and positive predictor of user satisfaction from using the CRS. Elavarasen (2010) also claimed that EaseOfUse has a positive impact on Attitude toward using e-Health Companion. Aldosari et al. (2018) also stated that EaseOfUse has a positive effect on nurses' acceptance toward using EMR. Kim et al. (2015) stated that Effort Expectency (EE) positively influenced the attitude to use a MEMR. EE has similar concept with UseF. Therefore, the following hypothesis is proposed about VHC:

**H4:** Perceived ease of use (EaseOfUse) will have a significant and positive effect on Attitude.

**4.5 The Relationship between Access and Attitude**
Accessibility of a system (Access) is defined as a capability of the system to be obtained, entered, affected, or comprehend. (The Concise Oxford Dictionary, 1982, p. 6). Several studies stated that perceived accessibility of information provided by a system is a more significant factor than the quality of the information in the system for adoption (O"Reilly, 1982; Culnan, 1984; Rice and Shook, 1988). Elavarasen (2010) claimed that Access has a positive impact on Intention toward using e-Health Companion. Therefore, the following hypothesis is proposed about VHC:

**H5:** Accessibility (Access) will have a significant and positive effect on Attitude.

**4.6 The Relationship between Comp and Attitude**
Rogers (1995, p. 224) defines the compatibility of a system (Comp) as the degree to which a new system is sensed to be coherent with the current values, previous experiences, and desires of possible adopters. Comp plays an important role in making end users feel positive about working with the system. Chen et al. (2002) and Wu and Wang (2005) stated that Comp is a more powerful factor than UseF. They considered Comp as a major driver in the online environment. Elavarasen (2010) also claimed that Comp has a positive influence on Attitude toward using e-Health Companion. Kuo et al. (2013) also stated that the compatibility of MEMRS with nurses' work practices enhances nurses' willingness to use MEMRS. The absence of integration into healthcare practices causes an electronic health system to fail (M. Gagnon et al., 2016). Therefore, the following hypothesis is proposed about VHC:

**H6:** Compatibility (Comp) will have a significant and positive effect on Attitude.

**4.7 The Relationship between PrivConc and Attitude**
The privacy and security concerns of information of patients are the barrier to adoption of e-Health services (Ponemon Institute LLC, 2014). It is critical to protect the privacy of medical information of patients. A system that is not protected against hackers and unauthorized users' access disappoints its end users. The following hypothesis examines the negative effect of privacy concern (PrivConc) on Attitude toward using VHC:

**H7:** Privacy concern about patients' personal information (PrivConc) will have a significant and negative effect on Attitude.

**4.8 The Mediating Effect of Attitude**
A variable is taken into account as a mediator when it makes the indirect effect through which the independent focal variable can influence the criterion variable of interest (Baron and Kenny, 1986). Attitude is included as a mediator in the proposed models by Davis (1989), Venkatesh and Davis (2000) and Venkatesh et al. (2008). In our study, we examine the mediating effect of attitude between our independent variables (UseF, EaseOfUse, Access, Comp, and PrivConc) and the criterion variable of interest (Intention). The following hypotheses examine the mediating effect of attitude:

**H8:** Attitude mediates the relationship between perceived usefulness (UseF) and Intention.

**H9:** Attitude mediates the relationship between perceived ease of use (EaseOfUse) and Intention.

**H10:** Attitude mediates the relationship between accessibility (Access) and Intention.

**H11:** Attitude mediates the relationship between compatibility (Comp) and Intention.

**H12:** Attitude mediates the relationship between Privacy concern about patients' personal information (PrivConc) and Intention.

## 5. Research Methodology

We applied an empirical study that is quantitative originally. A self-directed questionnaire was designed based on the variables to be analyzed. The items per each variable were measured on a 5-point Likert scale using scales from "Strongly disagree" to "Strongly agree" using cross-sectional data. The items are shown in Table 10. Moreover, some snapshots of VHC were included in the questionnaire. We also explained how the system works. Therefore, we expect that the respondents acquired enough knowledge about VHC.

The population of the current study covered all government hospitals and government health clinics listed in the ministry of health Malaysia's directory. As it was not practical to cover all of them in the sample, only 350 government hospitals and government health clinics were selected based on simple random sampling. Consequently, every single item in the population gained same probability of being selected. Besides, once an item in the population is selected, it is impossible that the item is selected again in the sample size (Kumar R., 2008).

350 hard copies of the questionnaire were distributed among them. We asked nurses to fill in the questionnaire. The questionnaires were delivered physically to them by post. Only 75 of 350 questionnaires were returned. During data screening, three questionnaires were not entertained in data analysis due to incompleteness. The respondents did not answer to the majority of questions. Moreover, two questionnaires were dropped due to lack of engagement of respondents because they selected the same answer for all questions. Finally, 70 questionnaires were used for data analysis. Table 2 shows the number of questionnaires used.

We studied two groups of multiple regression analysis. First, we looked at the relationship between independent variables (UseF, EaseOfUse, Access, Comp, and PrivConc) and Attitude. The sample size for this regression was calculated using G*Power analysis (Faul et al., 2009) at 80 percent with effect size 0.35 (Cohen, 1998). The minimum power needed in social and behavioral science research is commonly 0.8 (Rasoolimanesh et al. 2015). Since there were a total of five predictors, the minimum sample size was estimated at 43. In the second regression, we looked at the relationship between UseF and Attitude with the dependent variable (Intention). We also calculated the sample size by G*Power using the same criteria, but this time we considered the number of predictors as two. The suggested sample size was at 31. We considered 43 as the number of a needed sample size to fulfill the minimum requirement of sample size for both regressions. The number of collected questionnaires is 70. Therefore, data were collected beyond the minimum needed sample size which is 43. We can safely conclude that our sample size is acceptable for each group of MRA.

A methodology which is based on mix method of multiple regression analysis (MRA) and artificial neural networks (ANNs) applied. In this regard, after testing the goodness of measures and analyzing the effect of independent variables on the perspective dependent variable using the MRA, ANNs were applied. This method enables extra verification of the results provided by the MRA. Besides, for testing the mediating effect of Attitude on the relationship between the independent variables (UseF, EaseOfUse, Access, Comp, and PrivConc) and Intention, we applied the test for indirect effect suggested by Preacher and Hayes (2004, 2008) and bootstrapped the sampling distribution of the indirect effect, which is applicable for simple and multiple mediator models (Hair et al., 2014).

Table 2. Number of Questionnaires

| | | |
|---|---|---|
| Number of Questionnaires Distributed | 350 | |
| Number of Questionnaires Collected | 75 | |
| Response Rate | 75/350 x 100= 21% | |
| Number of Questionnaires Used for Analysis | 70 Rate =70/350 x 100= 20% | |

| Abbreviation | Items | Code |
|---|---|---|
| **UseF** (Adapted from Davis 1989) | • VHC can improve my efficiency in reminding the parents about their children vaccination uptakes.<br>• VHC can be helpful in managing vaccination plan for children. | Usef1<br><br>Usef4 |
| **Comp** (Adapted from Moore and Benbasat 1991) | • VHC is comptable to the way I like doing things.<br>• VHC fits into my work style. | Comp2<br>Comp3 |
| **EaseOfUse** (Adapted from Moore and Benbasat 1991) | • My interaction with VHC is clear and understandable.<br>• It is simple for me to become skilled at using VHC's components. | EaseOfUse2<br>EaseOfUse3 |
| **PrivConc** (Adapted from Elavarasen 2010) | • VHC cannot protect the parents and their children' personal information against illegal access.<br>• The parents and their children' personal information within VHC is vulnerable to fraudulent activity. | PrivConc1<br><br>PrivConc4 |
| **Access** (Adapted from Elavarasen 2010) | • VHC can provide children's vaccination information more accessible.<br>• It is easy for me to access to VHC. | Access2<br><br>Access3 |
| **Attitude** (Adapted from Taylor and Todd 1995b) | • Using VHC is a wise idea.<br>• Using VHC is a good idea. | Attitude3<br>Attitude4 |
| **Intention** (Adapted from Venkatesh et al. 2003) | • I predict I will use VHC at my hospital/clinic.<br>• I will use VHC in the future at my hospital/clinic. | Intention1<br>Intention2 |

Table 10. The Items per each Variable

*Note. Access1, Access4, EaseOfUse1, EaseOfUse4, Usef2, Usef3, PrivConc2, PrivConc3, Comp1, Attitude1, Attitude2 and Intention3 were dropped due to low item loadings or loading on several constructs.*

## 6. Analysis and Results

### 6.1 Demographic Statistic

A demographic profile of the participants is presented in Table 3. Most participants (55.8 percent) were aged less than 34 years. The majority of the participants are female (70 percent). The highest level of education is University degree (47.1 percent). Most participants have not applied any vaccination and management system (88.6 percent).

Table 3, Demographic Profile and Descriptive Statistics

*Note.* Standard Deviation: SD; [a] one-way Analysis of Variance;

## 6.2 Goodness of Measures

The goodness of measures shows whether an instrument or scale used is reliable and valid. They illustrate to what extent an instrument is accurately and consistently measuring a specific concept (reliability), and whether the instrument is indeed measuring the concept that it is supposed to measure (validity; Sekaran & Bougie, 2011). In the current study, the measurement items are evaluated for construct (factorial structure), convergent, and discriminant validity. The items which belong to one variable are also assessed for the reliability via their internal consistency of measures.

### 6.2.1 Construct Validity

Construct validity concerns the degree to which the test items measure the construct (variable) they were aimed to measure. Researchers frequently use factor analytic techniques to evaluate the construct validity of the scores acquired from an instrument (McCoach, 2002). Factor analysis offers an extensive group of methods and mathematical tactics for figuring out the latent variable structure of observed variables (Nunnally, 1978).

In this study, an exploratory factor analysis (EFA) with an orthogonal rotation of Varimax was applied to assess the construct validity of the instrument. We applied a principal component analysis (PCA) on the set of 26 items, which belong to the proposed model's constructs. The result of this analysis is shown in Table 4. The Anti-image correlation matrix diagonals are above 0.5 which show sampling adequacy. The analysis extracted a seven-factor solution, which is explaining 86.611% of the entire variance. The Kaiser–Meyer–Olkin measure of sampling adequacy was 0.744 showing a middling level in line with Kaiser and Rice (1974). The Bartlett's test for sphericity was significant ($x^2$= 525.872, p <

|  | Frequency | Frequency Percent | Intention Mean (SD) | Test and Sig |
|---|---|---|---|---|
| **Age categories** |  |  |  |  |
| 18-24 years | 16 | 22.9 | 4.28 | $F(5,64)$=1.55 |
| 25-34 years | 23 | 32.9 | 3.85 | $P_a$=0.19 |
| 35-44 years | 14 | 20 | 3.71 |  |
| 45-54 years | 11 | 15.7 | 4.04 |  |
| 55-64 years | 4 | 5.7 | 4 |  |
| 65 years or above | 2 | 2.9 | 4 |  |
| **Highest education** |  |  |  |  |
| Secondary school | 8 | 11.4 | 4.12 | $F(3,66)$=0.38 |
| Diploma | 16 | 22.9 | 3.84 | $P_a$=0.76 |
| University degree | 33 | 47.1 | 3.97 |  |
| Master | 13 | 18.6 | 4 |  |
| PhD | 0 | 0 | 0 |  |
| **Gender of Respondents** |  |  |  |  |
| Male | 21 | 30 | 3.95 | $F(1,68)$=0.011 |
| Female | 49 | 70 | 3.97 | $P_a$=0.92 |
| **Already applied any vaccination reminder and management system within your health center.** |  |  |  |  |
| No | 62 | 88.6 | 3.97 | $F(1,68)$=0.19 |
| Yes | 8 | 11.4 | 3.87 | $P_a$=0.67 |

2.0983E-61). Factors 1 to 7 were named as Access, EaseOfUse, UseF, PrivConc, Comp, Attitude, and Intention respectively.

Based on the rotated component matrix, of the 26 items, 12 items were dropped because of low item loadings of less than 0.50 (Hair, Black, Babin, and Anderson, 2010) and also due to cross loading. The outcomes drawn from the EFA showed that the items loaded on the constructs that they were supposed to represent. Consequently, the adequacy and construct validity of the items are obtained.

Table 4. Results of the EFA.

|  | 1 | 2 | 3 | 4 | 5 | 6 | 7 | Communality |
|---|---|---|---|---|---|---|---|---|
| Access3 | **.881** | .184 | .218 | .056 | .134 | .012 | .196 | .880 |
| Access2 | **.858** | .078 | .280 | .010 | .060 | .122 | .202 | .872 |
| EaseOfUse3 | .091 | **.925** | .071 | .003 | .190 | .013 | .106 | .897 |
| EaseOfUse2 | .147 | **.861** | .163 | .068 | .227 | .225 | .004 | .916 |
| Usef1 | .228 | .131 | **.874** | -.136 | .044 | .098 | .130 | .876 |
| Usef4 | .372 | .152 | **.769** | .084 | .198 | -.045 | .266 | .871 |
| PrivConc4 | .135 | .071 | -.005 | **.899** | .008 | .198 | -.010 | .874 |
| PrivConc1 | -.082 | -.004 | -.039 | **.884** | .185 | -.135 | .182 | .733 |
| Comp3 | .018 | .209 | .133 | .021 | **.917** | .149 | .121 | .872 |
| Comp2 | .229 | .368 | .071 | .298 | **.748** | .144 | -.100 | .940 |
| Attitude4 | .060 | .227 | -.032 | -.032 | .151 | **.827** | .266 | .764 |
| Attitude3 | .140 | -.021 | .442 | .325 | .325 | **.578** | -.057 | .834 |
| Intention1 | .317 | .092 | .207 | .158 | .072 | .155 | **.817** | .879 |
| Intention2 | .429 | .031 | .362 | .056 | -.058 | .334 | **.547** | .918 |
| Eigenvalue | 5.201 | 2.114 | 1.647 | 0.991 | 0.931 | 0.672 | 0.568 |  |
| % Variance | 37.153 | 15.101 | 11.767 | 7.081 | 6.648 | 4.801 | 4.059 |  |

*Note. Access= Accessibility; EaseOfUse= Perceived ease of use; Usef= Perceived usefulness; PrivConc= Privacy concern; Comp= Compatibility of the system; Attitude= Attitude toward using; Intention= Intention to use. The items in boldface indicate that these items fall under the same construct.*
*Access1, Access4, EaseOfUse1, EaseOfUse4, Usef2, Usef3, PrivConc2, PrivConc3, Comp1, Attitude1, Attitude2 and Intention3 were dropped due to low item loadings or loading on several constructs.*

### 6.2.2 Discriminant Validity

Discriminant validity refers to the extent to which measures of distinct constructs are relatively distinctive, that their correlation values are neither an absolute value of 0 nor 1 (Campbell & Fiske, 1959). Constructs which might be similar would have high correlations while unrelated constructs would result in low correlations. High correlation values of .90 and above is a sign of considerable collinearity among the constructs, indicating that the constructs are not distinct from each other (Hair et al., 2010).

A correlation analysis applied to the seven factors and the result is provided in Table 5. All constructs are not highly correlated with each other as their coefficients are less than 0.90, thus implying that the constructs are distinct from each other. Therefore, we can conclude that discriminant validity has been established.

Table 5. Means and Intercorrelations.

|  | Access | EaseOfUse | Usef | PrivConc | Comp | Attitude | Intention |
|---|---|---|---|---|---|---|---|
| Access | 1 |  |  |  |  |  |  |
| EaseOfUse | 0.325** | 1 |  |  |  |  |  |
| Usef | 0.607** | 0.314** | 1 |  |  |  |  |
| PrivConc | 0.102 | 0.097 | 0.026 | 1 |  |  |  |
| Comp | 0.293* | 0.540** | 0.313** | 0.291* | 1 |  |  |
| Attitude | 0.327** | 0.364** | 0.342** | 0.231 | 0.474** | 1 |  |
| Intention | 0.630** | 0.251* | 0.558** | 0.178 | 0.201 | 0.465** | 1 |

*Note Access= Accessibility of the system; EaseOfUse= Perceived ease of use; Usef= Perceived usefulness; PrivConc= Privacy concern; Comp= Compatibility of the system; Attitude= Attitude toward using; Intention= Intention to use*
*$*p<0.05$ and $**p<0.01$.*

### 6.2.3 Convergent Validity

In addition to the construct validity test using factor analysis (between scales), another factor analysis was performed, but this time using the within scale approach to examine for convergent validity. Convergent validity refers to all items measuring a construct loading on a single construct (Campbell and Fiske, 1959). This validity is established when all items measuring a construct all fall into one factor as theorized. Convergent validity was performed via a within-factor analysis (Hair, Black, Babin, & Anderson, 2010). All of the seven factors showed unidimensionality in the sense that each set of items demonstrating their intended concept loaded on only one construct. Therefore, we can conclude that convergent validity has been established.

### 6.2.4 Reliability Statistic

Cronbach's alpha is the fundamental formula used to decide the reliability established on internal consistency (Kim and Cha, 2002). Referring to Table 6, the Cronbach's alpha values for UseF, Comp, EaseOfUse, PrivConc, Access, and Intention is higher than 0.7. The minimum standard of 0.7 is set by Nunnally to pass this test (1978). Therefore, the internal consistency of the measurement for these variables is strong. However, the Cronbach's alpha for Attitude is less than 0.7. Anyhow, we keep this variable. The justification for this decision is that the number of items in the scale influences the Cronbach's alpha value. When there are a small number of items in the scale (less than 10), the Cronbach's alpha value can be quite small (Vaske et al. 2016). As the discussion in the construct validity section, we dropped two items from Attitude. Consequently, the number of items for Attitude is small and it influences the Cronbach's Alpha value of this construct.

Table 6, Result of Reliability Analysis

| Variables | Abbreviation | Number of Items | Cronbach's Alpha |
|---|---|---|---|
| Perceived usefulness | UseF | 2 | 0.858 |
| Compatibility | Comp | 2 | 0.843 |
| Perceived ease of use | EaseOfUse | 2 | 0.888 |
| Privacy concern | PrivConc | 2 | 0.795 |
| Accessibility | Access | 2 | 0.895 |
| Attitude toward using | Attitude | 2 | 0.500 |
| Intention to use | Intention | 2 | 0.753 |

### 6.3 Hypotheses Testing

Multiple Regression Analysis (MRA) examines the relationship between a single dependent variable (continuous) and several independent variables (continuous or even nominal). We studied the relationships between two groups of continuous variables using two groups of MRA. For the first regression, we examined the relationship between independent variables (UseF, EaseOfUse, Access, Comp, and PrivConc) and Attitude. In the second regression, we examined the relationship between UseF and Attitude with the dependent variable (Intention).

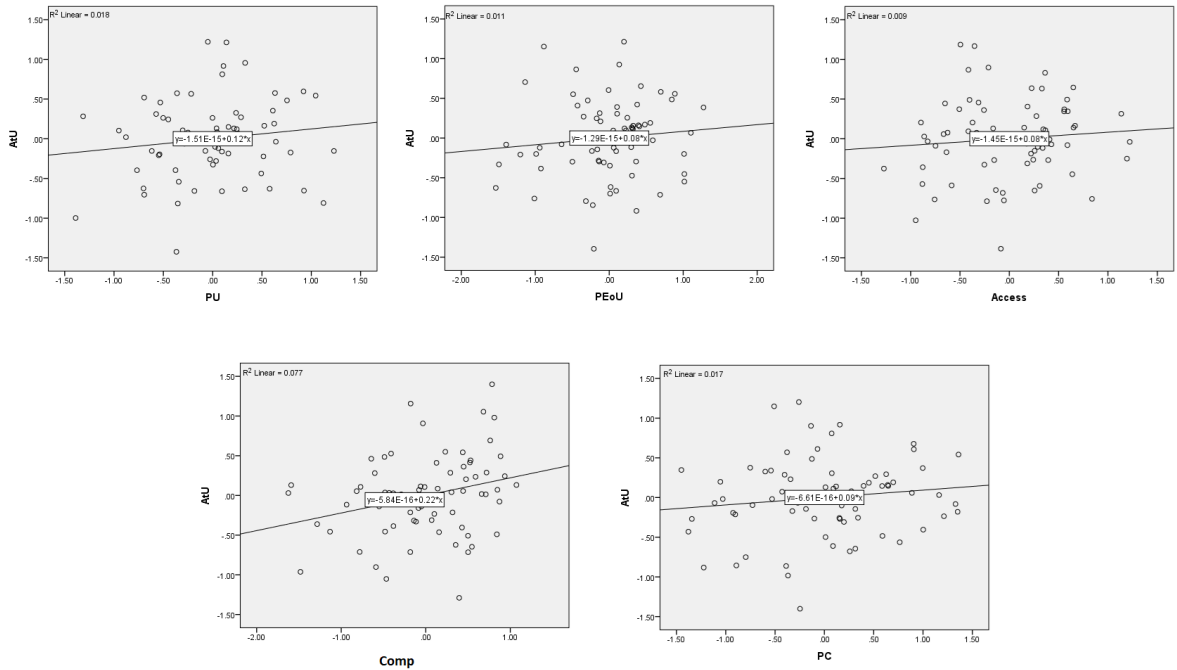

**Fig 9**, Partial regression plots for Attitude
*Note. AtU: Attitude, PU: UseF, PEoU: EaseOfUse, PC: PrivConc*

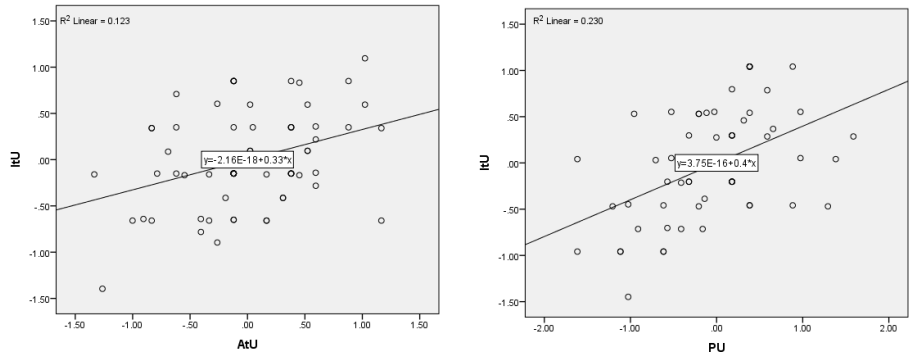

**Fig 10**, Partial regression plots for Intention
*Note. ItU: Intention, PU: UseF, AtU: Attitude*

Before we interpret a regression analysis, we need to fulfill several assumptions, which are 1) normality, 2) normality of the error terms, 3) linearity, 4) constant variance – homoscedasticity, 5) multicollinearity, and 6) independence of the error term – autocorrelation 7) outliers. We could assume that both regressions passed normality assumption because the histogram plots showed not much abnormality. Both histogram plots obtained mean value close to 0 and standard deviation value close to 1. Regarding the second assumption, we referred to Normal P-P Plots of both regressions. In both plots, the points were close to the diagonal line. Therefore, we passed this assumption. For linearity assumption, we looked at partial plots for each relationship (as shown in Figure 9 and 10). We passed this assumption because we could draw a straight line on each plot. The fourth assumption was that the variance must be constant (Homoscedasticity) as opposed to not constant (Heteroscedasticity). Heteroscedasticity is detected when we realize a consistent pattern when we plot the studentized residual (SRESID) against the predicted value of Y (ZPRED). In both plots, we could not find a consistent pattern. Therefore, we passed this assumption for both regressions. For assumption fifth, we checked that the independent variables are not highly correlated with each other. All VIF values for each variable for both regressions were less than 5. Moreover, the conditional index values for both regressions were less than 30. Therefore, we could assume that there was no collinearity problem in our data. In independence of the error term – autocorrelation assumption, we want to make sure that each predicated value is independent and it is not related to any other prediction such as time. To do that, we looked at Durbin-Watson values for both regressions. Durbin-Watson values for both regressions were between 1.5-2.5. Therefore, we passed assumption sixth. The last assumption is about outliers. They are values that are extremely large and powerful that they can influence the results of the regression. The threshold for both regressions was set at +- 3 standard deviations. We passed this assumption because we could not detect any outliers in our data for both regressions. After passing all the assumptions, we can interpret the results of both regressions.

In first MRA, the relationship between Attitude and UseF, EaseOfUse, Access, Comp, and PrivConc are examined (as shown in table 7). These five factors together contribute to 29% of the variance in Attitude. Among five factors, Comp (Std. Beta =0.305) only has a significant and positive impact on Attitude (p < 0.05). The other factors are not significant.

In the second regression, the relationship between Intention and the predictor variables (Attitude and UseF) were examined. Table 8 shows that all proposed hypotheses are supported. As stated by R square of 0.397, 40 percent of the variance in Intention is described by the predictor variables. Intention has two significant predictor variables that are UseF (Std. Beta = 0.452) and Attitude (Std. Beta = 0.310). The relationships are significant and positive at 0.01 level.

The bootstrapping analysis (table 11) showed that there was a significant indirect effect of UseF, EaseOfUse, Access, Comp, PrivaCon on Intention through the mediating effect of Attitude. As mentioned by Preacher and Hayes (2008), when the indirect effect does not straddle a 0 in between Boot CI's LL and UL, there is a mediation. Therefore, we can conclude that the mediating effect of Attitude is statistically significant.

**Table 7,** Multiple regression results of Attitude
$R^2 =0.295$, *p < 0.05. Std. error of the estimate: 0.515

| Hypothesis | Abbreviation | Unstd. Beta | Std. Beta | Std. Error | t-value | Decision |
|---|---|---|---|---|---|---|
| H2 | UseF | 0.124 | 0.148 | 0.113 | 1.095 | Not supported |
| H4 | EaseOfUse | 0.085 | 0.109 | 0.099 | 0.851 | Not supported |
| H5 | Access | 0.083 | 0.101 | 0.111 | 0.745 | Not supported |
| H6 | Comp | 0.222 | 0.305 | 0.096 | 2.306* | Supported |
| H7 | PrivConc | 0.093 | 0.117 | 0.088 | 1.061 | Not supported |

**Table 8,** Multiple regression results of Intention
$R^2 =0.397$, **p<0.01, Std. error of the estimate: 0.490

| Hypothesis | Abbreviation | Unstd. Beta | Std. Beta | Std. Error | t-value | Decision |
|---|---|---|---|---|---|---|
| H1 | Attitude | 0.326 | 0.310 | 0.106 | 3.072** | Supported |
| H3 | UseF | 0.398 | 0.452 | 0.089 | 4.479** | Supported |

**Table 11,** Mediating effects of Attitude

| Hypothesis | Beta | Boot CI LL | Boot CI UL | Decision |
|---|---|---|---|---|
| Usef → Attitude → Intention | 0.093 | 0.027 | 0.217 | Supported |
| EaseOfUse → Attitude → Intention | 0.128 | 0.038 | 0.243 | Supported |
| Access → Attitude → Intention | 0.082 | 0.024 | 0.177 | Supported |
| Comp → Attitude → Intention | 0.173 | 0.077 | 0.322 | Supported |
| PrivConc → Attitude → Intention | 0.086 | 0.014 | 0.188 | Supported |

*Note. 95 % confidence interval*

**6.4 Multi-Analytical Approach of MRA and ANN**

We mixed Multiple Regression Analysis (MRA) and ANN to apply a multi-analytical approach. Therefore, first, MRA was used to check the overall research model and detects significant predictors. Second, significant predictors are used as inputs to the ANN model to determine the average relative importance of each predictor variable. ANN is one of the most important artificial intelligence techniques (Liébana-Cabanillas et al. 2017). Traditional linear statistical techniques such as MRA are only capable of detecting the linear relationships. Therefore, they are suitable for hypotheses testing, which is based on regression line's slope. However, MRA is incapable of helping decision makers to make a precise decision due to weak prediction accuracy. Therefore, we applied ANN since this computational model can provide higher prediction accuracy. Consequently, ANN can out-perform MRA and other conventional techniques (Liébana-Cabanillas et al. 2017; Tan et al., 2014; Leong et al. 2013).

In first MRA, only one predictor is statistically significant. Therefore, we did not apply the multi-analytical approach. For second MRA, we have built 10 ANNs in SPSS 23. Significant variables of the second MRA (Usef and Attitude) are used as the input units for ANN. We selected Intention as the dependent variable. To avoid over-fitting, a 10-fold cross validation was done. Therefore, 90% of the data was used for network training, and the last 10% was used for testing to measure the prediction accuracy of the trained network (Liébana-Cabanillas et al. 2017). Initially, the architectures of the ANNs were designed automatically by SPSS. However, due to low prediction accuracy, we used two hidden layers. Moreover, the number of neurons in each hidden layer is two. Besides, we selected Sigmoid as an activation function for hidden layers and output layer. Batch is selected as a type of training technique due to our small sample size. As a measure of the predictive accuracy of the model, the Root Mean Square of Errors (RMSEs) was used. Moreover, the normalized importance in the sensitivity analysis is calculated per each input node of ANN. The relative powers of the causal relationships were tested based on the normalized importance in the sensitivity analysis (Tan et al., 2014).

Figures 11 shows the RMSE of training and testing data for Intention. The ANN number 3 has the higher prediction accuracy due to lowest RMSE for testing. Figure 12 represents the structure of ANN number 3 for Intention.

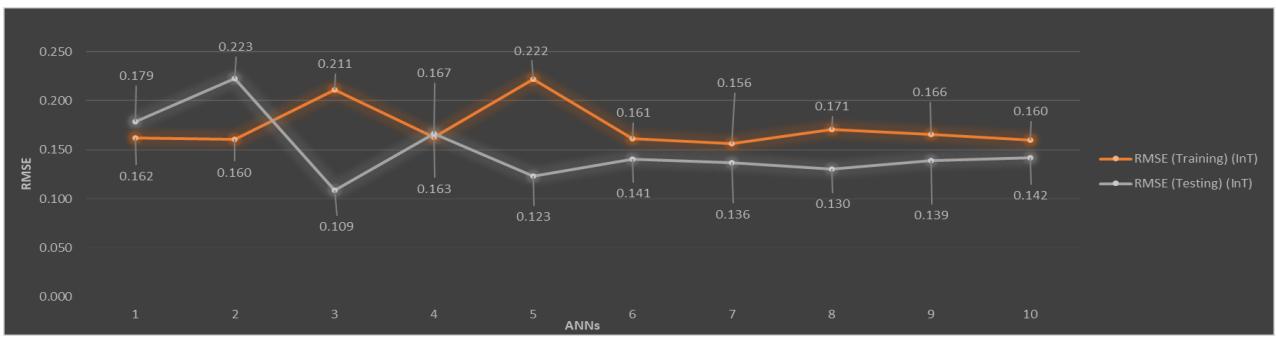

**Fig. 11.** The RMSEs of training and testing data for Intention
*Note: Int: Intention, ANNs: Artificial Neural Networks, RMSEs: The Root Mean Square of Errors*

Referring to Figures 11, the highest value for RMSE for training and testing data records are 0.222 and 0.223 respectively. Therefore, the ANN model is very trustworthy to measure the relationships between predictors (Usef, Attitude) and output (Intention) (Leong, 2013; Tan, 2014; Liébana-Cabanillas et al., 2017).

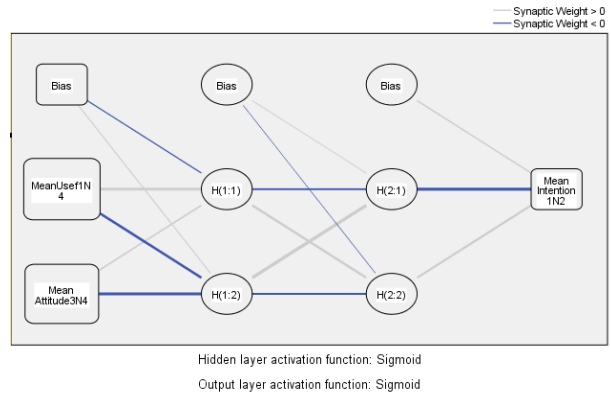

**Fig 12,** The structure of ANN number 3 for Intention
*Note. MeanUsef1N4: UseF, MeanAttitude3N4: Attitude, MeanIntention1N2: Intention*

The importance of each independent variable is a measure of how properly the expected values of the ANN model alters with response to different values in the independent variable (Leong et al., 2013). In the sensitivity analysis, the normalized importance of the input variables was calculated by dividing the importance of each input variable with the largest importance value in percentage form. Since only significant linear factors from the MRA were used as the input units of the ANN models, only linear relationships were identified (Tan et al., 2014). Based on the presented ANNs, Usef (with average normalized importance = 100 %) is the most significant predictor of Intention, followed by Attitude (with average normalized importance = 69 %). The normalized importance of the two factors is shown in Figure 13.

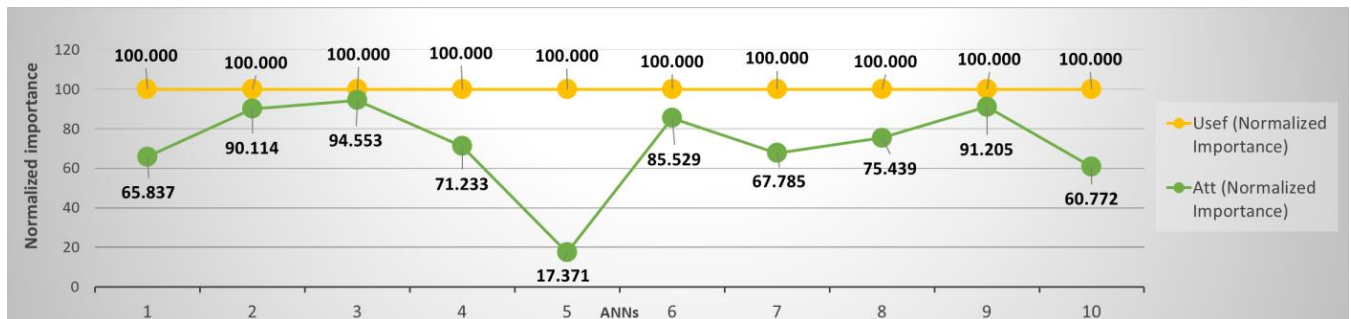

**Fig. 13**, The normalized importance of Usef and Attitude for Intention
*Note: ANNs: Artificial Neural Networks, Att: Attitude*

## 7. Discussion

The first and second objectives of the current study are to examine the relationship between Attitude and its predictor variables and also Intention and its predictor variables. Therefore, MRA is applied. In first MRA, Comp only had a significant and positive effect on Attitude among other factors. This result is consistent with Chen et al. (2002), Wu and Wang (2005) who stated that Comp is a more powerful factor than UseF in the online environment. Gagnon et al (2016) stated the importance of compatibility of an e-health system into healthcare practices in order to achieve a successful implication of that system. Kuo et al. (2013) also indicated that the compatibility of MEMRS with nurses' work practices improves nurses' willingness to use MEMRS. This result of our study shows that the compatibility of the system is the most influential and significant factor in point of view of health centers in order to give them positive feelings about working with the system. A system that highly fits in with the health centers' culture and user desires is highly compatible and easy to accept (Hsiao and Chen, 2015). The predictor variables of Attitude accounted for 29.5 % of the total explained variance in Attitude.

We could not prove the significant relationship between PrivCon and Attitude. The possible explanation for the insignificant relationship is due to that we do not record the sensitive information of the parents and their children on the system.

The relationships between UseF, EaseOfUse, Access and Attitude are also not supported. The health centers use a paper-based vaccination system, and majority of them did not applied any vaccination reminder and management system. Therefore, IT educational programs can be provided for nurses to improve their information technology literacy to help them recognizing the benefits of VHC (Kuo et al., 2013).

The result of the second MRA indicated that UseF had a significant and positive effect on Intention. Previous works on TAM model supported the relationship between UseF and Intention (Davis, 1989; Venkatesh and Davis, 2000; Venkatesh et al., 2008). Moreover, TAM and extended TAM proposed by Marie-Pierre G et al. (2014) confirmed this result. Marie-Pierre G et al. (2016) also supported this relationship. Aldosari et al. (2018) also stated that UseF had a positive influence on nurses' acceptance toward using EMR. Dünnebeil et al. (2012) also stated that UseF positively influenced behavioral intention to use Electronic Health Services (EHS). Wahyuni (2017) also indicated that UseF positively influenced intention to use e-health services. Kuo et al. (2013) also indicated that the Usef of MEMRS enhances nurses' intention to use MEMRS. Aldosari (2012) also stated that UseF of a PACS positively influenced intention to use PACS. Dunnebeil et al. (2012) also proved that UseF positively influenced the intention to use e-health applications. Esmaeilzadeh et al. (2015) studied the intention to use a CDSS and indicated PE as an influential factor. Kijsanayotin et al. (2009) also stated that PE influenced the intention to use medical IT. Hsiao and Chen (2015) also indicated that UseF of CCPG was critical factor influencing physicians' intention to use CCPG.

Moreover, Attitude had a significant and positive effect on Intention. This result is supported by Davis (1989) and Ajzen and Fishbein (1980). Owitlawakul et al. (2014) also stated that attitude toward using the Electronic Health Records for Nursing Education (EHRNE) software program was the influential factor on intention to use the EHRNE. Kijsanayotin et al. (2009) also indicated that Attitude influenced the intention to use medical IT. Kim et al. (2015) indicated that Attitude positively influenced the intention to use a MEMR. Hsiao and Chen (2015) also indicated that attitudes toward using CCPG was critical factor influencing physicians' intention to use CCPG. The predictor variables of Intention accounted for 39.7 % of the total explained variance in Intention. All predictor factors of Intention were significant. Therefore, ANN was applied for checking the significance of each significant variable detected in MRA. Interestingly, the trained ANN could estimate the significance of each predictor factor with less error compared to MRA. Based on the ANNs, UseF is the most significant predictor of Intention, followed by Attitude. It can be interpreted that health centers' intention to use the system is further influenced by their perceived usefulness of the system than their positive feelings about working with the system.

In order to achieve the third objective which is to examine the mediating effect of Attitude, a resampling strategy (bootstrapping) by Preacher and Hayes (2008) is applied. The bootstrapping analysis showed that there was a significant indirect effect of each independent variable, i.e., UseF, EaseOfUse, Access, Comp, PrivaCon on Intention through the mediating effect of Attitude. Davis (1989), Venkatesh and Davis (2000) and Venkatesh et al. (2008) supported the mediating effect of Attitude. To the best knowledge of the authors, the mediating effect of Attitude is proved to be significant for the first time in acceptance studies of EHR systems. It is important to consider the indirect effect and direct effect of independent variables on intention to use the system.

In order to achieve the last objective, A one-way analysis of variance (ANOVA) was conducted to determine whether there are any statistically significant difference in Intention scores accounting for the participants' demographic characteristics (age, education, gender and whether they already applied any vaccination reminder and management system). The assumptions of this test are met. However, we could not determine any statistically significant difference in Intention scores accounting for the participants' demographic characteristics. The results of one-way ANOVA are represented in Table 3.

## 8. Limitation of Study and Scope for Further Research

The first limitation is that the sample size may not be representative of the whole country since this research has been conducted only in big cities of Malaysia. Therefore, we recommend a bigger sample size to test the research model further. Another restriction is that this research is limited to study of only seven variables which are supported by the literature. For further study, we recommend to include other factors such as IT knowledge of nurses in the acceptance model. The current study only performed among nurses. Therefore, different staff members, such as physicians, technicians should be considered in order to get a better understanding about acceptance of VHC. We also recommend to consider the characteristic of health centers including linking structures between staff, size, and the amount of slack resources to enrich the research model further.

## 9. Conclusion

Comp is the only significant factor that influences Attitude. Therefore, it is critical for the software developers to ensure that the system fits into health centers' work style and fulfills their needs to make health centers feel positive about VHC. IT educational programs also should be provided for nurses to improve their information technology literacy to help them to identify the benefits of VHC. Moreover, that health centers' intention is further influenced by their perceived usefulness than their positive feelings. Therefore, firstly, the software developers need to ensure that the system improves the efficiency of health centers in reminding about vaccination uptakes, and is helpful in managing vaccination plan in order to gain the health centers' aim to use VHC. Secondly, they need to assure that VHC is a wise idea in point of view of the health centers. On the other hand, the mediating effect of Attitude is approved. Therefore, it is important to consider the indirect effects from independent variables (UseF, EaseOfUse, PrivConc, Comp and Access) on intention to use the system. We could not determine any statistically significant difference in Intention scores accounting for the participants' demographic characteristics. Therefore, the software developers do not need to consider the demographic characteristics of the end users during the development of the system.

**Abbreviations**

CRS: Clinical Reminder System; GO: General Optimism; CK: Computer Knowledge; CE: Computer Experience; IniU: Initial Usage; AU: Average Usage; SAT: User Satisfaction; SRU: Self-reported Usage; IU: Institutionalized Use; UTG: Usage Trajectory Group; DR: Demonstrability of the Results; CSE: Computer Self-Efficacy; PI: Personal Identity; ICT: Information and Communication Technology; SN: Social Norm; PN: Professional Norm; RC: Resistance to Change; IC: Information about the Change; EMR: Electronic Medical Record; EHRNE: Electronic Health Records for Nursing Education; MEMR: Mobile Electronic Medical Record; CCPG: Computerized Clinical Practice Guidelines; EHR: Electronic Health Record; EHS: Electronic Health Services; MEMRS: Mobile Electronic Medical Record System; PACS: Picture Archiving and Communication System; CDSS: Clinical Decision Support System; PE: Performance Expectancy; IT: Information Technology; EFA: Exploratory Factor Analysis; PCA: Principal Component Analysis; AtU: Attitude to Use; PU: Perceived Usefulness; PEoU: Perceived Ease of Use; PC: Privacy Concern; ItU: Intention to Use; AtU: Attitude to Use; SRESID: Studentized residual; ZPRED: Predicted value of Y; VIF: Variance Inflation Factors; Std: Standardized; Unstd: Unstandardized; SPSS: Statistical Package for the Social Sciences; RMSEs: Root Mean Square of Errors; ANOVA: Analysis of variance.

**Declarations**
**-Ethics Approval and Consent to Participate**
This study is permitted by school of computer sciences, USM. The participants voluntarily agreed to take part in this study.
**-Consent to Publish**
Not applicable.
**-Availability of Data and Materials**
The data analysed during the current study are available from the corresponding author on reasonable request.
**-Competing Interests**
The authors declare that they have no competing interests.
**-Funding**
The work reported in this paper is supported by Universiti Sains Malaysia's APEX Delivering Excellence 2012 Grant.
**-Acknowledgement**
The authors wish to thank Keng Siang Ooi for his invaluable assistance in making VHC available for this study.